\documentclass[letter]{aa}  

\usepackage{txfonts}
\usepackage{graphicx}   

\usepackage[colorlinks,linkcolor=blue,citecolor=blue]{hyperref}
\newcommand{\urm}{$\rm rad\,m^{-2}$}

\definecolor{Brown}{rgb}{0.647,0.165,0.165}

\let\upmu\muup

\begin{document} 

    \title{Detection of magnetic fields in the circumgalactic medium of nearby galaxies using Faraday rotation}


   \author{V. Heesen\inst{1}
          \and
          S.~P.~O'Sullivan\inst{2}
          \and
          M.~Br\"uggen\inst{1}
          \and
          A.~Basu\inst{3}
          \and
          R.~Beck\inst{4}
          \and
          A.~Seta\inst{5}
          \and
          E.~Carretti\inst{6}
          \and
          M.~G.~H.~Krause\inst{7}
          \and
          M.~Haverkorn\inst{8}
          \and
          S.~Hutschenreuter\inst{8}
          \and
          A.~Bracco\inst{9}
          \and 
          M.~Stein\inst{10}
          \and
          D.~J.~Bomans\inst{10}
          \and
          R.-J.~Dettmar\inst{10}
          \and 
          K.~T.~Chy\.zy\inst{11}
          \and
          G.~H.~Heald
          \inst{12}
          \and
          R.~Paladino\inst{6}
          \and
          C.~Horellou\inst{13}
          }

   \institute{Hamburg University, Hamburger Sternwarte, Gojenbergsweg 112, 21029 Hamburg, Germany\\
              \email{volker.heesen@hs.uni-hamburg.de}
         \and
         School of Physical Sciences and Centre for Astrophysics \& Relativity, Dublin City University, Glasnevin, D09 W6Y4, Ireland
         \and
         Th\"uringer Landessternwarte, Sternwarte 5, 07778 Tautenburg, Germany 
         \and 
         Max-Planck Institut f\"ur Radioastronomie, Auf dem H\"ugel 69, 53121 Bonn, Germany
         \and
         Research School of Astronomy and Astrophysics, Australian National University, Canberra, ACT 2611, Australia
         \and
         INAF, Istituto di Radioastronomia, Via Gobetti 101, I-40129 Bologna, Italy
         \and
         Centre for Astrophysics Research, University of Hertfordshire, College Lane, Hatfield AL10 9AB, UK
         \and
         Department of Astrophysics/IMAPP, Radboud University, PO Box 9010, 6500 GL Nijmegen, The Netherlands
         \and
         Laboratoire de Physique École Normale Supérieure (LPENS), 24 rue Lhomond, 75005, Paris, France
         \and
         Ruhr University Bochum, Faculty of Physics and Astronomy, Astronomical Institute, 44780 Bochum, Germany
         \and
         Astronomical Observatory, Jagiellonian University, ul. Orla 171, 30-244 Krak\'ow, Poland
         \and
         CSIRO Space and Astronomy, PO Box 1130, Bentley WA 6102, Australia
         \and
         Dept of Space, Earth and Environoment, Chalmers University of Technology, Onsala Space Observatory, 43992 Onsala, Sweden 
       }

   \date{Received 30 January 2023 / Accepted 10 February 2023}
   
 
  \abstract
   {The existence of magnetic fields in the circumgalactic medium (CGM) is largely unconstrained. Their detection is important as magnetic fields can have a significant impact on the evolution of the CGM, and, in turn, the fields can serve as tracers for dynamical processes in the CGM.}
   {Using the Faraday rotation of polarised background sources, we aim to detect a possible excess of the rotation measure in the surrounding area of nearby galaxies.}
   {We used 2,461 residual rotation measures (RRMs) observed with the LOw Frequency ARray (LOFAR), where the foreground contribution from the Milky Way is subtracted. The RRMs were then studied around a subset of 183 nearby galaxies that was selected by apparent $B$-band magnitude.}
   {We find that, in general, the RRMs show no significant excess for small impact parameters (i.e.\ the perpendicular distance to the line of sight). However, if we only consider galaxies at higher inclination angles and sightlines that pass close to the minor axis of the galaxies, we find significant excess at impact parameters of less than 100~kpc. The excess in |RRM| is $3.7$~\urm\ with an uncertainty between $\pm 0.9$~\urm\ and $\pm 1.3$~\urm\ depending on the statistical properties of the background ($2.8\sigma$--$4.1\sigma$). With electron densities of $\sim$$10^{-4}~\rm cm^{-3}$, this suggests magnetic field strengths of a few tenths of a microgauss.}
   {Our results suggest a slow decrease in the magnetic field strength with distance from the galactic disc, as expected if the CGM is magnetised by galactic winds and outflows.}



\titlerunning{Detection of magnetic fields in the circumgalactic medium with Faraday rotation}
\authorrunning{V.~Heesen et al.}

   \maketitle
%

\section{Introduction}

The evolution of galaxies is regulated by the accretion and expulsion of matter that is related to feedback by either star formation or active galactic nuclei. In recent years it has become clear that galaxies are surrounded by large reservoirs of a diffuse, hot, and tenuous gas now referred to as the circumgalactic medium \citep[CGM;][]{tumlinson_17a,faucher_giguere_23a}. While the CGM comprises multi-phase gas, it is so tenuous that it can be studied via absorption lines in the spectra of bright background quasars that show the presence of highly ionised species \citep{lehner_15a,werk_16a}; a tentative detection in X-ray emission was reported by \citet{das_20a}. Theoretical work suggests that the CGM may have substantial pressure support from cosmic rays \citep{buck_20a}. Magnetic fields are another pressure component that needs to be considered when determining the dynamics and structure of the CGM \citep{van_de_voort_21a}.

Two possible scenarios have been put forward to explain how the CGM might get magnetised: the magnetic fields are either generated in situ by turbulence via the small-scale dynamo \citep{pakmor_17a} or they are transported from the galaxies via galactic winds \citep{peroux_20a}. Both models suggest that higher magnetic field strengths are found near the minor axis of galaxies as outflows, either stellar-  \citep{thomas_22a} or active-galactic-nucleus-driven \citep{pillepich_21a}, and are driven most easily along the minor axis, which both transports magnetic fields and generates turbulence. The processes that magnetise the CGM will also influence its structure. The small-scale dynamo preferentially generates turbulent magnetic fields; via shear, these magnetic fields can be converted into anisotropic fields \citep{fletcher_11a}. If shear exists in the CGM on sufficiently large scales, one can expect anisotropic random fields on scales of tens of kiloparsecs. Similarly, if the gaseous halo of a galaxy hosts a dynamo, one would have large-scale magnetic field configurations \citep{moss_10a}. Lastly, models of galactic winds show a helical magnetic field structure that results in a large-scale ordering \citep{zirakashvili_96a,henriksen_16a,thomas_22a}; such large-scale helical magnetic fields have indeed been found in galaxy haloes  \citep{mora_19b,stein_20a}.

Because magnetic fields in the CGM are expected to be weak, they can be detected best via the Faraday rotation of polarised background sources, rather than directly via synchrotron emission due to the effects of spectral ageing \citep{miskolczi_19a}. Faraday rotation can be described by a wavelength-dependent shift in the polarisation angle,
\begin{equation}
    \varphi(\lambda) = \varphi_0 + {\rm RM} \lambda^2,
    \label{eq:faraday_rotation}
\end{equation}
where $\varphi$ and $\varphi_0$ are the observed and intrinsic polarisation angles, respectively, RM is the rotation measure, and $\lambda$ is the wavelength. This definition of the RM is only valid for a synchrotron source behind the Faraday rotating medium, whereas mixed media are described by a Faraday spectrum with various Faraday rotating components, called Faraday complexity \citep{brentjens_05a}. Neglecting Faraday complexity, the RM is the line-of-sight (LoS) integral,\begin{equation}
    {\rm RM}= 0.81 \int_{\rm LoS} \left (\frac{n_e}{\rm cm^{-3}}\right ) \left ( \frac{B_\parallel}{\upmu\rm G} \right ) \left ( \frac {{\rm d}r}{\rm pc} \right )~{\rm rad\,m^{-2}},
    \label{eq:rotation_measure}
\end{equation}
where $n_{\rm e}$ is the electron number density and  $B_{\parallel}$ the strength of the magnetic field component along the LoS.
One way to study the effect of the CGM is to use background sources where a Mg\,{\sc ii} quasar absorption line is detected, which is indicative of the presence of ionised gas \citep{kacprzak_08a}. It has been shown that Mg\,{\sc ii} absorbers have higher absolute RM values, |RM|, than the control sample at frequencies of 5~GHz \citep{bernet_08a,bernet_10a} but possibly not at lower frequencies \citep{bernet_12a}. At lower frequencies, the effect of Faraday rotation becomes stronger (see Eq.~\ref{eq:faraday_rotation}), so an inhomogeneous distribution of the RM can lead to increased depolarisation \citep{kim_16a}. However, at $1.4$~GHz, \citet{farnes_14a} were able to show that, even at such a low frequency, the excess in |RM| can still be seen. \citet{farnes_14a} also find that the excess is only found in sources where the sightlines to the quasar and the polarised emission are in good agreement. More recently, however, these results have been questioned by \citet{lan_20a} and \citet{amaral_2021a}, who do not find a correlation of |RM| with the number of foreground galaxies; much larger and precise datasets are required to study this further \citep{basu_18a, shah_21a}.

Studies thus far have concentrated on galaxies at moderate redshifts ($z\sim 0.5$) as they are more prone to show strong absorption lines. However, this has a downside because the measurable RM signal is suppressed by a factor of $(1+z)^{-2}$. It is, however, an intriguing possibility that ordinary nearby galaxies are also surrounded by a magnetised CGM that can potentially be detected with this method. This has now become feasible as low-frequency radio continuum polarimetry leads to a higher precision, such as by using the LOw Frequency ARray \citep[LOFAR;][]{vanHaarlem_13a}. Using data from the LOFAR Two-metre Sky Survey \cite[LoTSS;][]{shimwell_17a,shimwell_19a}, \cite{osullivan_23a} find 2,461 polarised sources at 144~MHz. Compared with current surveys at $1.4$~GHz, the source density per sky area is about five times lower due to depolarisation, but the values of the RM can be much more precise because of the strong wavelength dependence. In this Letter we probe the excess |RM| as a function of the impact parameter of nearby galaxies with an apparent optical $B$-band magnitude of $m_B<12.5~\rm mag$ drawn from the Palomar sample of \citet{ho_97a}. We find a marginally significant ($\gtrsim$2$\sigma$) excess of |RM| at impact parameters less than 100~kpc, which indicates the presence of a magnetised CGM.

\begin{figure}
    \centering
    \includegraphics[width=0.6\linewidth]{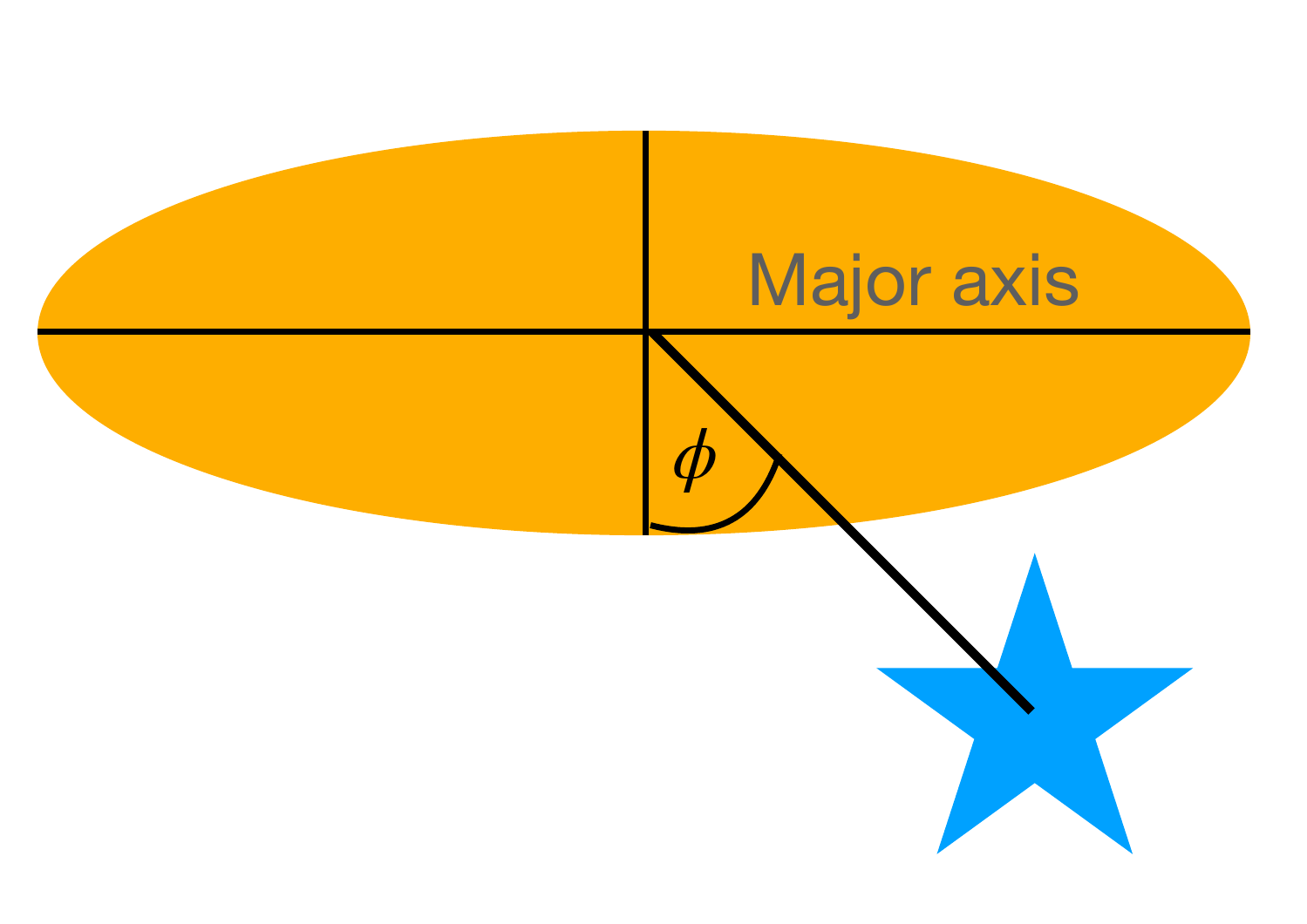}
    \caption{Azimuthal angle, $\phi$, of the polarised background source with respect to the minor axis. Sources near the minor axis have $|\phi|<45\degr$, whereas sources near the major axis have $|\phi| > 45\degr$.}
    \label{fig:sketch}
\end{figure}
\begin{figure*}
    \centering
    \includegraphics[width=0.49\linewidth]{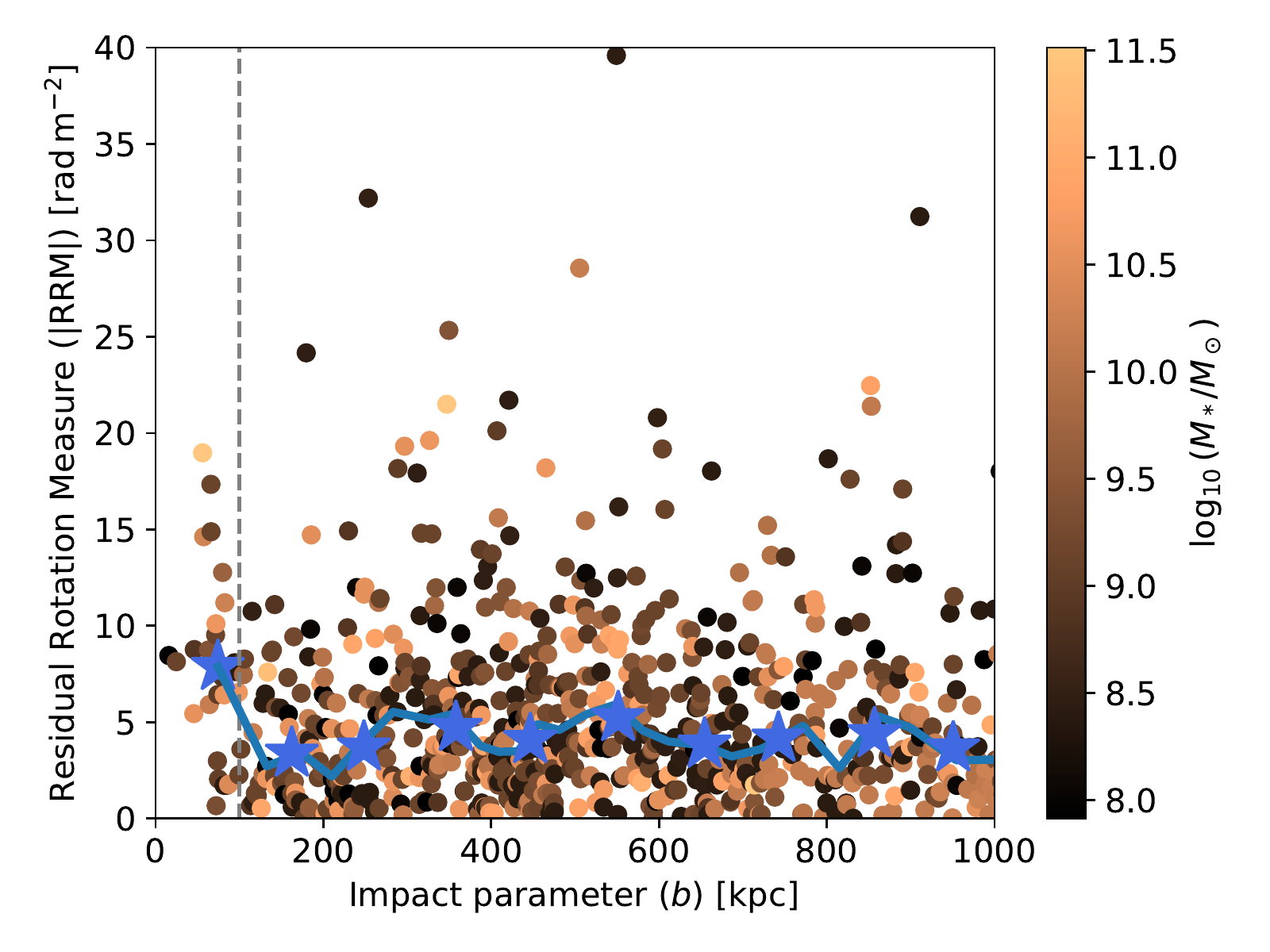}
    \includegraphics[width=0.49\linewidth]{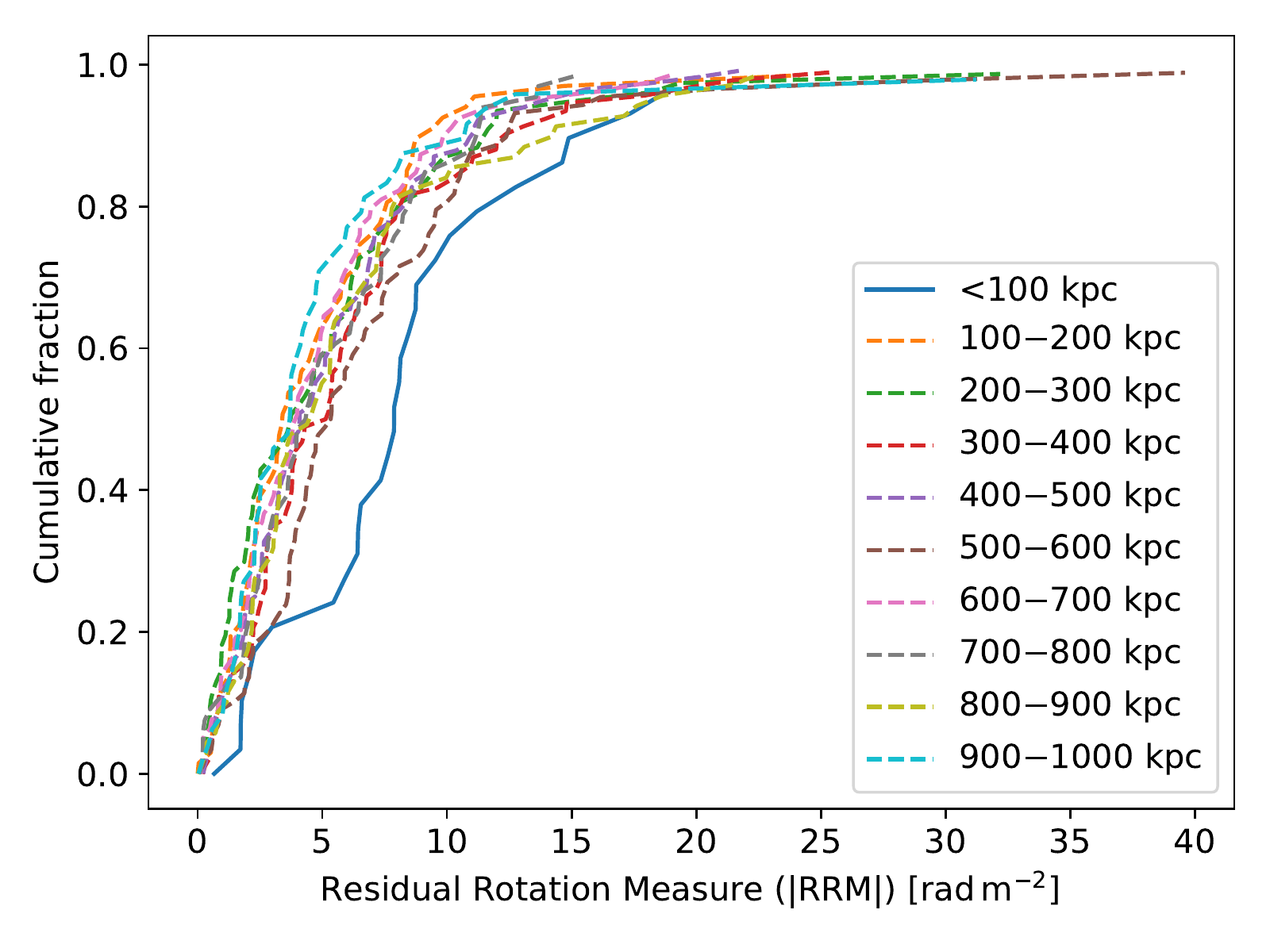}
    \caption{Excess |RRM| at small impact parameters to the foreground galaxies. {\it Left panel:} |RRM| as a function of impact parameter for inclined galaxies   ($i\geq 55\degr$) where the background polarised source lies near the minor axis ($|\phi|<45\degr$). Data points are coloured according to the stellar mass of the foreground galaxy. Blue stars show the median |RRM| binned in 100 kpc intervals; the standard deviation is $0.57\pm 0.13$~\urm\ ($b>100$~kpc) and $1.1$~\urm\ ($b<100$~kpc). The solid blue line shows the running median of 29 binned data points, and the vertical dashed line is at $b=100$~kpc. {\it Right panel:} Cumulative distribution function of the |RRM| for sightlines binned into 100 kpc intervals. The solid blue line shows sightlines with $b<100$~kpc, whereas the dashed coloured lines show them for $b>100$~kpc.}
    \label{fig:rm_all}
\end{figure*}


\begin{figure}
    \centering
    \includegraphics[width=\linewidth]{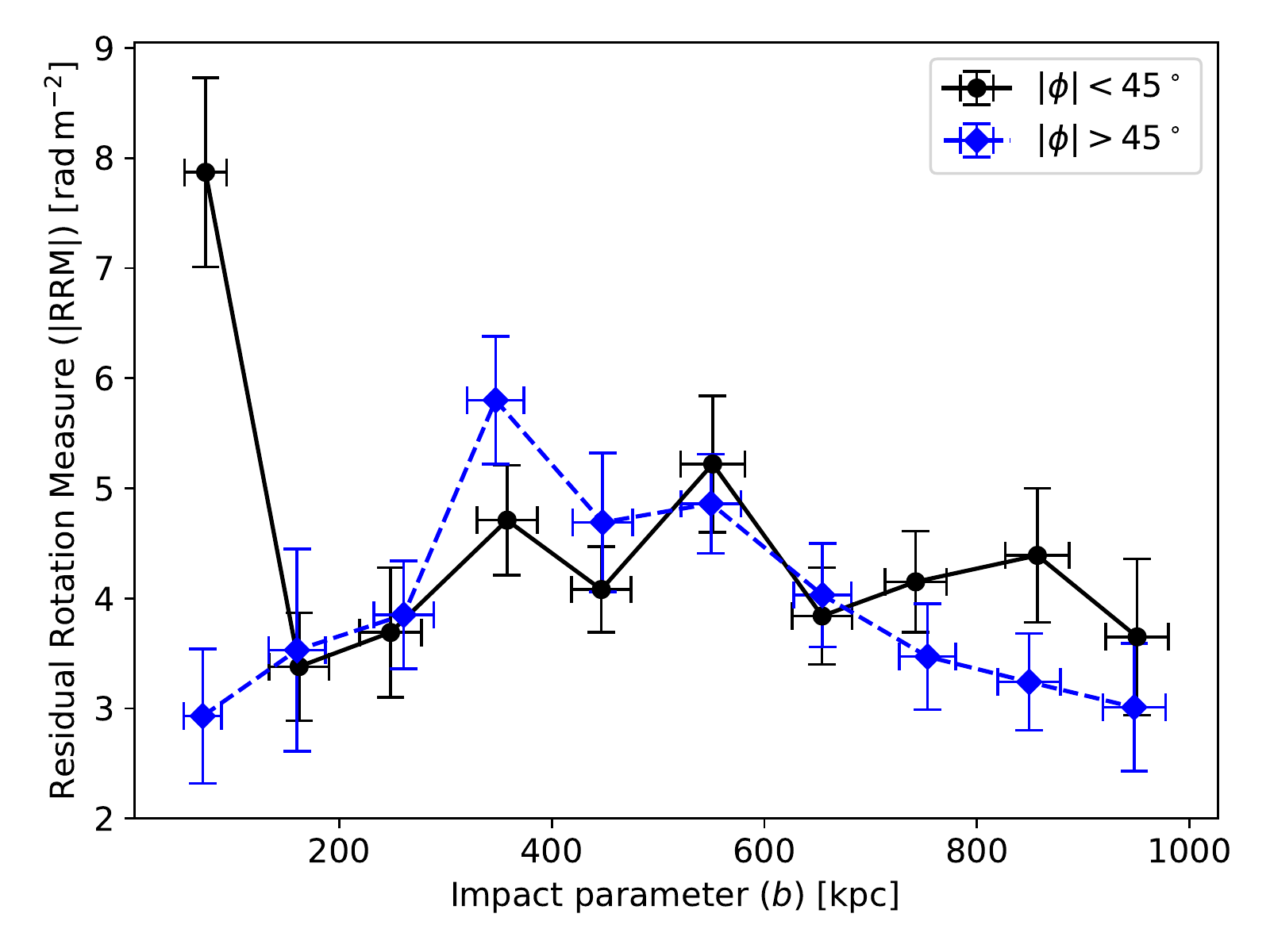}
    \caption{|RRM| as a function of the impact parameter for different azimuthal angles relative to the disc minor axis for inclined galaxies. The filled black circles with solid lines are the measured |RRM| values close to the minor axis of the disc (i.e. $|\phi|<45\degr$), and the dashed line with filled diamonds give the measured |RRM| values near the major axis of the disc (i.e. $|\phi|>45\degr$). The error bars of the impact parameter give the standard deviation and those of |RRM| the standard error of the mean.}
    \label{fig:quadrant}
\end{figure}

\begin{figure}
    \centering
    \includegraphics[width=\linewidth]{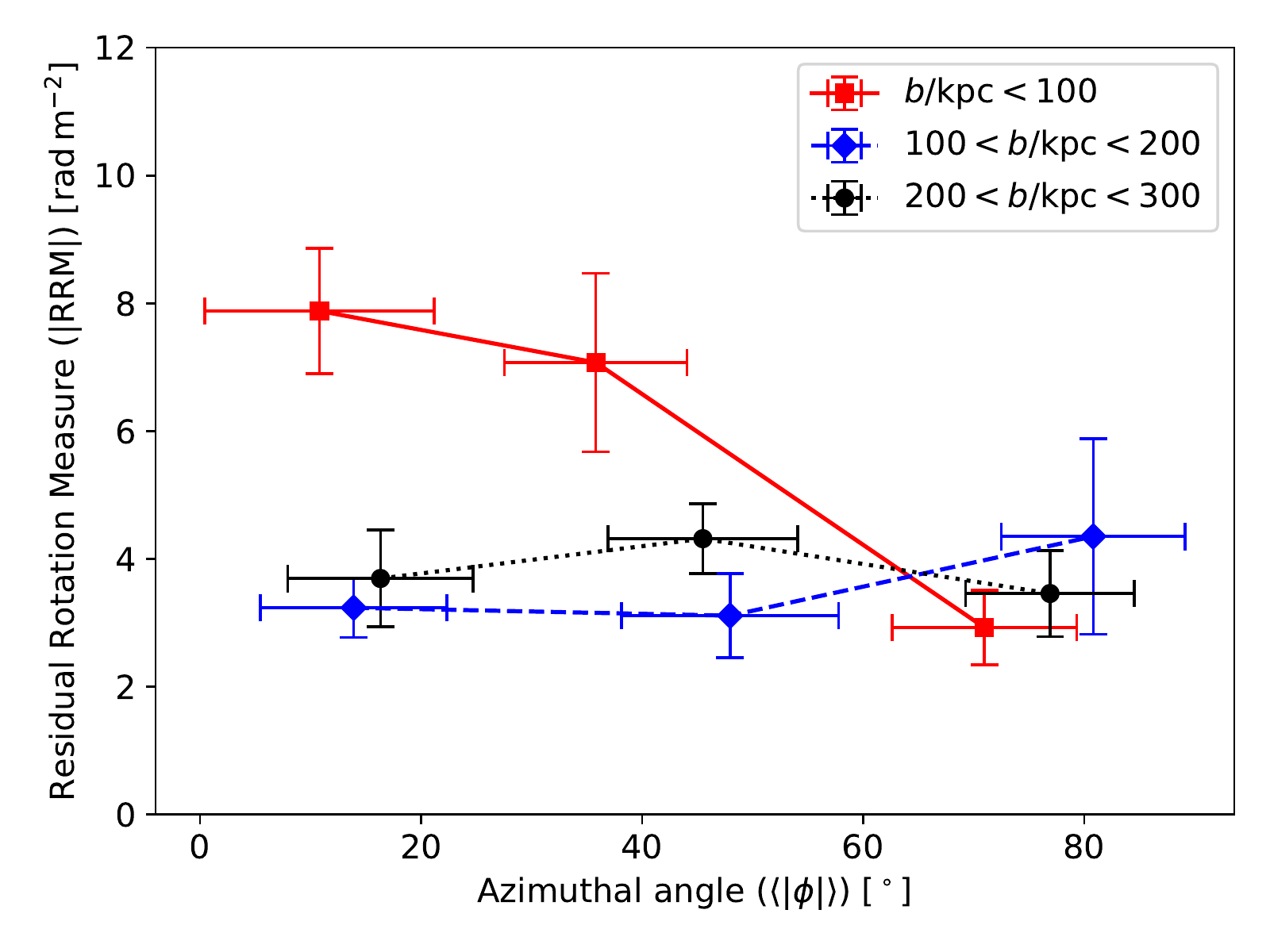}
    \caption{Dependence of the |RRM| on the azimuthal angle relative to the disc minor axis for inclined galaxies. Three azimuthal angular bins within $0\degr<|\phi|<30\degr$, $30\degr < |\phi| < 60\degr$, and $60<|\phi|<90\degr$ of the minor axis are defined for each of the three bins in terms of the impact parameter: $b < 100~\rm kpc$ (red squares), $100~{\rm kpc} < b < 200~\rm kpc$ (blue diamonds), and $200~{\rm kpc}<b<300~\rm kpc$ (black circles). The error bars of the angles are the standard deviation of the mean of the azimuthal angles within that bin.}
    \label{fig:azimuthal}
\end{figure}

\section{Data and methodology}
\label{s:data}

We used a catalogue of high-precision Faraday RMs from \citet{osullivan_23a}, who present 2,461 extragalactic sources. These RMs have a median uncertainty of only $0.06$~\urm, which is the random uncertainty, although a higher systematic error of $0.3$~\urm\ is possible after ionospheric correction. Comparisons with centimetre-wavelength RMs indicate minimal amounts of Faraday complexity \citep{livingston_21a} in the LoTSS detections, meaning that they are well suited for studying Faraday rotation in the foreground. These data were taken at a frequency of 144~MHz and at an angular resolution of $20\arcsec$ using the LOFAR High-Band Antennas \citep{vanHaarlem_13a}. The sources are distributed over an area of 5,720~$\rm deg^2$ on the sky using data from the LoTSS Data Release 2 \citep[LoTSS-DR2;][]{shimwell_22a} observed at declinations greater than $0\degr$ with a frequency range between 120 and 168~MHz. 
The Milky Way contribution to the RM, the Galactic rotation measure (GRM), is subtracted, leaving us with the residual rotation measure (RRM):  $\rm RRM=RM-GRM$, using the GRM model from \citet{hutschenreuter_20a}. In particular, we used the mean value of the reconstructed Faraday sky (i.e.~the GRM) constrained by RM values with a typical separation of $1\degr$ \citep{oppermann_12a} and also constrained by the emission measure data from \textit{Planck} \citep{adam_16a}. The LoTSS RM values are not included in this GRM reconstruction. We subtracted the GRM from the LoTSS RM at the location of each source using the interpolated value of the four nearest neighbours in the GRM map. We also tested averaging the GRM value over a disc of $1\degr$ in diameter as well as using the GRM map, which is not constrained by the emission measure data.\ In both cases, this had an insignificant effect on the results.

We selected galaxies from the Palomar sample of \citet{ho_97a}, 183 of which overlapped with our covered area. This sample is magnitude selected, so it contains the brightest nearby galaxies where our subset has a median distance of 18~Mpc ($3.2$--109~Mpc). We obtained coordinates in right ascension ($\alpha$) and declination ($\delta$) from the NASA Extragalactic Database (NED). We also obtained redshift-independent measurements of the distance to the galaxies from NED, giving preference to distances obtained with Cepheids or the method using the tip of the red giant branch. If that was not available, we used distances from the Tully--Fisher relation or the method of brightest stars. We obtained the positions and inclination angles from the HyperLEDA data base \citep{makarov_14a}. We used 2 $\upmu$m fluxes from the Two Micron All Sky Survey \citep{huchra_12a} to calculate stellar masses.


For each polarised background source, we then calculated the impact parameters as the perpendicular distance between the LoS and the galaxies via
\begin{equation}
    b_{jk} = 17.5\sqrt{\cos^2(\delta_k)(\alpha_j-\alpha_k)^2+(\delta_j-\delta_k)^2} d_k~\rm kpc,
\end{equation}
where $j$ specifies the background source and $k$ the foreground galaxy. The units for $\alpha_{j,k}$ and $\delta_{j,k}$ are degrees and the distance, $d_k$, is in units of Mpc. The minimum impact parameter, $b,$ for each background source was then retained. We also calculated the azimuthal angle for each background source with respect to the minor axis of the galaxy, where an azimuthal angle of $0\degr$ means that the source is lying along the minor axis of the galaxy (see Fig.~\ref{fig:sketch}). Any galaxies from the Local Group at $d<1~\rm Mpc$ were excluded as their signal, having an apparent scale of several degrees, would be subtracted together with the Milky Way foreground.

\section{Results}
\label{s:results}


First, we analysed the RRM for all galaxies, finding no excess of |RRM| at small impact parameters (not shown here). Next, we analysed the RRM for all inclined galaxies (i.e.\ with an inclination angle $i\geq 55\degr$). If we select only galaxies where the background source lies near the minor axis with $|\phi|<45\degr$, as shown in Fig.~\ref{fig:rm_all} (left panel), we find an excess of |RRM| for impact parameters $b<100$~kpc, where the median |RRM| is $7.8\pm 0.9$~\urm; we estimated the error by bootstrapping. For impact parameters $100<b/{\rm kpc}<1000,$ the distribution of the binned median in 100 kpc intervals, |RRM|, is nearly flat and can be fitted with a constant value of $4.1\pm 0.2$~\urm. Hence, the |RRM| excess would be $3.7\pm 0.9$~\urm\ at a $4.1\sigma$ significance. This assumes that all sightlines at $b>100$~kpc trace the background and contain no signal. A more conservative estimate of the background uncertainty is to calculate the running median with a binning size of 29 data points, which is the number of sightlines at $b<100$~kpc. The standard deviation of these median |RRM| values is $1.0$~\urm, which means our background |RRM| estimate is $4.1\pm 1.0$~\urm. Thus, the excess |RRM| at small impact parameters is $3.7\pm 1.3$~\urm, equating to a $2.8\sigma$ significance.

The fact that the sightlines with $b<100$~kpc show an excess of |RRM| can also be seen in the cumulative distribution function shown in Fig.~\ref{fig:rm_all} (right panel). A two-sample Kolmogorov--Smirnov test shows that the distribution of the |RRM| for $b<100$~kpc has less than a 3\% probability of agreeing with any of the other 100~kpc bins, with an average probability of 0.5\%. Also, we note that the median photometric redshift for sources at $b<100$~kpc, $z=0.56\pm 0.41$,  is smaller than for sources at $b>100$~kpc, for which it is $0.69\pm 0.43$. Hence, the |RRM| excess cannot be explained by the boosting of the intrinsic source RM with higher redshift as the effect is less than 10\% \citep{basu_18a}. There is no difference between group and non-group members \citep{garcia_93a}.  A similar trend of |RRM| with the impact parameter is also seen from analytical models of thermal electron density and magnetic fields \citep[see Fig.~10 in][]{shah_21a}. The standard deviation for the |RRM| of all sightlines is $5.5$~\urm. Both values are in good agreement with the upper limit of $\approx$6~\urm\ for the extragalactic contribution \citep{schnitzeler_10a,vacca_16a}, suggesting that the origin of the scatter is the intrinsic RMs of the sources and the contribution from cosmic web filaments \citep{carretti_23a}, and a possible contribution from the Galactic foreground on scales of less than $1\degr$ \citep{sun_09a}. 

If we select galaxies where the background source lies near the major axis with $|\phi|>45\degr$, the excess |RRM| vanishes. This can be seen in Fig.~\ref{fig:quadrant}, where we show the |RRM| dependence on the azimuthal angle relative to the disc minor axis. Values of |RRM| near the minor axis show an excess at small impact parameters in comparison with  sources near the major axis. The difference disappears at larger impact parameters because the polar quadrants ($|\phi|<45\degr$) have a much steeper radial decline in |RRM|. By comparison, the radial profile in the plane of the disc is rather flat and agrees with the polar one at $b>200$~kpc. A similar effect is presented in Fig.~\ref{fig:azimuthal}, which shows the azimuthal dependence for three radial bins of impact parameters. At small impact parameters, there is an azimuthal gradient in |RRM| from the poles down to the plane of the disc, with an |RRM| about two times higher. This could be taken as evidence that the |RRM| excess at low impact parameters $\lesssim$100~kpc originates from the bipolar regions aligned with the minor axes of the galaxies. The dependence on the azimuthal angle has been observed in quasi-stellar-object absorption line measurements of Mg\,{\sc ii} \citep{bouche_12a}, where a bimodal distribution along the major and minor axis points to either large gaseous discs or galactic winds. Similarly, \citet{bordoloi_11a} find that only inclined galaxies ($i>50\degr$) show an excess of Mg\,{\sc ii} absorption along the minor axis. 





\section{Discussion}
\label{sec:discussion}

For the first time, our data show that an excess of |RRM| along sightlines around nearby galaxies can be detected along the minor axis of inclined galaxies for impact parameters of less than 100~kpc. The dependence on azimuthal angle is in agreement with quasi-stellar-object absorption line measurements that trace the presence of ionised gas. Also, the median mass of our galaxies with sightlines $b<100$~kpc is $M_\star=10^{9.1\pm 0.9}~\rm M_\sun$. Simulations show that biconical winds form in such galaxies with total masses between $10^{10}$ and $10^{11}~M_\sun$ (assuming a 10\% baryonic to total mass ratio); for lower masses the winds are spherical, whereas for higher masses they do not reach beyond the virial radius \citep[$\ge$150~kpc;][]{jacob_18a}. Hence, we argue that this effect is real and that the excess can be attributed to the CGM of the observed galaxies. The observed excess of $3.7\pm 1.3$~\urm\ is much smaller than the $24\pm 6$~\urm\ reported by \citet{farnes_14a}  in Mg\,{\sc ii} absorbers and even smaller than the $\approx$140~\urm\ of \citet{bernet_08a}. This is certainly an effect of  the insensitivity to large |RM| values, $\gtrsim$$50$~\urm, at LOFAR frequencies \citep{heald_09a}.


With an assumption of electron density, $n_e$, we can thus calculate the strength of the LoS component of the ordered magnetic field, $B_\parallel$, from Eq.~\eqref{eq:rotation_measure}. The current best estimate of the electron density in the CGM of the Milky Way is $\approx$$10^{-4}~\rm cm^{-3}$ at radii $\approx$50--100~kpc \citep[see Fig.~10 in][]{donahue_22a}. With a LoS length of 100~kpc, we obtain an average strength of the LoS magnetic field of $\approx$$0.5$~$\upmu$G. The energy density of this field component is $U_{\rm B}\approx 10^{-14}~\rm erg\,cm^{-3}$, in good agreement with the thermal energy density of the hot CGM  ($T\approx 10^6$~K), meaning that we find a plasma beta of $\beta\approx 1$. We note that this is an upper limit because the magnetic field strength may be higher since magnetic field reversals can lower the RM. Assuming a random distribution of the magnetic field with a coherence length of $l_c\approx 30$~kpc \citep[as in the intra-cluster medium;][]{subramanian_06a}, the actual field strength would be a factor of $\sqrt{L/l_c}\approx 2$ higher. We note that the contribution from turbulent fields in the CGM on scales smaller than the projected size of the background polarised emission can also, in part, contribute to an excess of |RRM| \citep{basu_18a}. We have ignored such contributions 
as this would 
require additional information on the nature of the frequency-dependent Faraday depolarisation.
Using data from the Very Large Array Sky Survey \citep{condon_98a}, 
\cite{AntonNilsson2016} find a decrease in the amount of polarised sources 
near galaxies and interpreted this as evidence for depolarisation by galactic halos. 



The magnetic field strength in the CGM is about one-tenth of the strength of the ordered magnetic field in the discs of galaxies, which points to a rather slow decrease with distance, $r$, from the galactic disc. Observations of nearby galaxies show for the total magnetic field $B=B_0\exp(-r/r_0)$, where $B_0\approx 10~\upmu$G and $r_0\approx 10$~kpc \citep{beck_15a}. For the ordered field, significantly higher scale heights are found \citep{krause_20a}. Assuming $r_0\approx 20$~kpc and $B_0\approx 5~\upmu$G, the ordered field strength would be the correct strength at $r=50$~kpc. Such high magnetic field strengths are indeed seen in simulations \citep{pakmor_20, thomas_22a}. One can also explain the higher polar RRM with a specific chimney structure, such as seen in simulations, where one would expect $\beta\approx 1$ from the shear between chimneys and halo gas \citep{rodgers_lee_19a, krause_21b}.

\section{Conclusions}
\label{sec:conclusions}



We used observations of the RM\ of linearly polarised background sources with LOFAR to detect the magnetic field in the CGM surrounding nearby galaxies. We chose the apparent luminosity-selected Palomar catalogue of nearby galaxies and the RM catalogue from \citet{osullivan_23a}. We subtracted the foreground RM from the Milky Way to obtain the RRM. Below we summarise our conclusions:

\begin{enumerate}
    \item If we consider only inclined galaxies (inclination angle larger than $55\degr$) and background sources within $45\degr$ of the minor axis, we find a moderate excess of |RRM| at impact parameters of $<$100~kpc of $3.7$~\urm\ with an uncertainty between $0.9$ and $1.3$~\urm\ ($2.8\sigma$--$4.1\sigma$).
    \item In contrast, no |RRM| excess is found for background sources within $45\degr$ of the major axis.
    \item The dependence on azimuthal angle is limited to impact parameters smaller than 100~kpc, showing the maximum extent of the magnetised CGM detectable with this method and instrument.     \item For an order-of-magnitude estimate, we assume an electron density of $10^{-4}~\rm cm^{-3}$ and a sightline length of 100~kpc, which provides us with an average strength of the LoS magnetic field of $0.5~\upmu$G.
\end{enumerate}

Our new results are in agreement with upper limits on |RRM| in the CGM in previous observations \citep{lan_20a} and suggest that magnetised bipolar galactic winds exist. Thus far, our method exploits only statistical properties of galaxies because for studies of individual galaxies the source density is too low. In the future, with more sensitive low-frequency telescopes, such as the upgraded LOFAR2.0 and the Square Kilometre Array, individual galaxies may come within reach, allowing the magnetic field structure to be studied in detail. For instance, the opening angle of the outflow cone or a change of the sign of the RRM across either major or minor axes, which would provide hints to the large-scale field structure, could be investigated. A complementary approach is to use high-frequency data such as from MeerKAT and the Australian Square Kilometre Array Pathfinder.


\begin{acknowledgement}
We thank the anonymous referee for an insightful report that helped to improve the paper. We wish to express our gratitude to Kofi Rogmann for aiding us in the data analysis. We also thank Olaf Wucknitz for carefully reading the manuscript. This paper is based (in part) on data obtained with the International LOFAR Telescope (ILT). LOFAR \citep{vanHaarlem_13a} is the Low Frequency Array designed and constructed by ASTRON. It has observing, data processing, and data storage facilities in several countries, that are owned by various parties (each with their own funding sources), and that are collectively operated by the ILT foundation under a joint scientific policy. The ILT resources have benefitted from the following recent major funding sources: CNRS-INSU, Observatoire de Paris and Universit\'e d'Orl\'eans, France; BMBF, MIWF-NRW, MPG, Germany; Science Foundation Ireland (SFI), Department of Business, Enterprise and Innovation (DBEI), Ireland; NWO, The Netherlands; The Science and Technology Facilities Council, UK; Ministry of Science and Higher Education, Poland.

MB acknowledges funding by the Deutsche Forschungsgemeinschaft (DFG, German Research Foundation) under Germany’s Excellence Strategy – EXC 2121 ‘Quantum Universe’ – 390833306. MH and SH acknowledge funding from the European Research Council (ERC) under the European Union's Horizon 2020 research and innovation programme (grant agreement No 772663). AB acknowledges support from the European Research Council through the Advanced Grant MIST (FP7/2017-2022, No.742719). RJD acknowledges funding from the German Science Foundation DFG, within the Collaborative Research Center SFB1491 "Cosmic Interacting Matters - From Source to Signal".
This research has made use of the NASA/IPAC Extragalactic Database (NED),
which is operated by the Jet Propulsion Laboratory, California Institute of Technology,
under contract with the National Aeronautics and Space Administration.
This work made use of the {\sc SciPy} project \href{https://scipy.org}{https://scipy.org}. 

\end{acknowledgement}

%
%

\bibliographystyle{aa}
\bibliography{review} 



    \label{fig:stellar}




\end{document}